\documentclass[nofootinbib,aps,twocolumn,10pt,floatfix,secnumarabic]{revtex4}

\usepackage{graphicx}
\usepackage{amsmath}
\usepackage{amssymb}
\usepackage{citesort}
\usepackage{afterpage}
\usepackage{alltt}

\begin{document}

\title{
LevelScheme: A level scheme drawing and scientific figure preparation
\\ system for Mathematica }

\author{M. A. Caprio}
\affiliation{Center for Theoretical Physics, Sloane Physics Laboratory, 
Yale University, New Haven, Connecticut 06520-8120, USA}

\date{\today}

\maketitle

\onecolumngrid

\noindent\rule{\hsize}{1pt}

~

\noindent\textbf{Abstract}

~

LevelScheme is a scientific figure preparation system for Mathematica.
The main emphasis is upon the construction of level schemes, or level
energy diagrams, as used in nuclear, atomic, molecular, and hadronic
physics.  LevelScheme also provides a general infrastructure for
the preparation of publication-quality figures, including support for multipanel
and inset plotting, customizable tick mark generation, and various
drawing and labeling tasks.  Coupled with Mathematica's plotting
functions and powerful programming language, LevelScheme provides a
flexible system for the creation of figures combining diagrams,
mathematical plots, and data plots.

~

\noindent\textbf{Program Summary}

~

\noindent\textit{Title of program:} LevelScheme

\noindent\textit{Catalogue identifier:} ADVZ

\noindent\textit{Program summary URL:} http://cpc.cs.qub.ac.uk/summaries/ADVZ

\noindent\textit{Program available from:} CPC Program Library, Queen's University of Belfast, N. Ireland

\noindent\textit{Operating systems:} Any which supports Mathematica;
tested under Microsoft Windows XP, Macintosh OS X, and Linux

\noindent\textit{Programming language used:} Mathematica 4

\noindent\textit{Number of bytes in distributed program, including test code and documentation:} 3\,051\,807

\noindent\textit{Distribution format:} tar.gz

\noindent\textit{Nature of problem:} Creation of level scheme
diagrams.  Creation of publication-quality multipart figures
incorporating diagrams and plots.

\noindent\textit{Method of solution:} A set of Mathematica 
packages has been developed, providing a library of level scheme
drawing objects, tools for figure construction and labeling, and
control code for producing the graphics.

~


~


~

\noindent\rule{\hsize}{1pt}

~

\twocolumngrid
\newpage  


\section{Introduction}
\label{secintro}
\makeatletter
\def\@makefntext#1{%
  \def\baselinestretch{1}%
  \reset@font\small
  \parindent 1em%
  ~\\[-15pt]
  \noindent
  #1\par
}%
\makeatother
\footnotetext{\textit{Published as} M.~A.~Caprio, Comput.~Phys.~Commun.~171 (2005) 107.}

LevelScheme is a scientific figure preparation system for
Mathematica~\cite{wolfram1999:mathematica-book-4,wolfram1999:mathematica-4}.
The main focus is upon the construction of level schemes, or level
energy diagrams, as used in several areas of physics, including
nuclear, atomic, molecular, and hadronic physics.  However, convenient
preparation of publication-quality figures requires a variety of
capabilities beyond those available with Mathematica's built-in
graphics display functions.  The LevelScheme system thus also provides support
for multipanel and inset plotting, customizable tick mark generation,
and general drawing and labeling tasks.  Coupled with Mathematica's
powerful programming language, plotting functions, and mathematical
typesetting capabilities, LevelScheme is a flexible system for the
creation of publication-quality figures.  Figures can combine data plots,
mathematical plots, and graphics generated by specialized packages
(\textit{e.g.}, Ref.~\cite{hahn2001:feynarts}) with diagrams
constructed using LevelScheme's drawing tools.

LevelScheme automates many of the tedious aspects of preparing a level
scheme, such as positioning transition arrows between levels or
placing text labels alongside the objects they label. The package
allows extensive manual fine tuning of the drawing appearance, text
formatting, and object positioning. It also includes specialized
features for creating several common types of level schemes
encountered in nuclear physics.  Note that there already exist
programs for drawing certain specific types of level schemes, such as
band structure diagrams or decay schemes (\textit{e.g.},
Refs.~\cite{radford1995:radware,dunford2003:ensdat}).

After a discussion of the general features of the LevelScheme system
(Sec.~\ref{secprinciples}), the major components of the software are
considered in greater depth separately
(Secs.~\ref{secobjects}--\ref{secinfra}).  Examples of the package's
graphical output are given in Sec.~\ref{secexamples}.  The more
technical details of using the LevelScheme package are addressed in
the documentation provided through the CPC Program Library, which
includes a tutorial discussion, a full reference guide, and extensive
examples of code for producing figures with LevelScheme.  Updates and
further information may be obtained through the LevelScheme home
page~\cite{levelschemehome,mathsourcecontribs}.

\section{Principles of implementation}
\label{secprinciples}

A few basic principles have guided the design of the LevelScheme
system.  One is that it should be possible for the user to make even
major formatting changes to a level scheme relatively quickly.
Objects are attached to each other (transition arrows are attached to
levels, labels are attached to levels or transition arrows,
\textit{etc.}), so that if one object is moved the rest follow
automatically.  Another principle is for objects to have reasonable
default properties, so that an unsophisticated level scheme can be
drawn with minimal attention to formatting features.  But the user
must then be given extensive flexibility in fine tuning formatting
details, to accomodate whatever special cases might arise.  This is
accomplished by making the more sophisticated formatting features
accessible through various optional arguments (``options'') for which
the user can specify values.  The user can specify the values of
options for individual objects, or the user can set new default values
of options for the whole scheme to control the formatting of many
objects at once.  Finally, attention has been paid to providing a
uniform user interface for all drawing objects, based upon a
consistent notation for the specification of properties for the
outline, fill, and text labels of objects.

The code for the LevelScheme system consists of essentially three
parts.  There is a library of functions made available to the user for
drawing the individual objects within a figure
(Sec.~\ref{secobjects}).  There are tools included to facilitate the
general aspects of figure preparation and layout, such as construction
of panels, axes, and tick marks (Sec.~\ref{sectools}).  And underlying
these is a general infrastructure for managing the drawing process
(option processing, coordinate system arithmetic, and a layered
drawing system) and the final display of the figure
(Sec.~\ref{secinfra}).

\section{Drawing object library}
\label{secobjects}

The drawing object library is the portion of the LevelScheme system
which is most visible to the user.  Some of the objects in this
library are specialized elements of level scheme diagrams,
while others are general purpose drawing shapes and labels.
\begin{figure*}
\begin{center}
\includegraphics*[width=0.68\hsize]{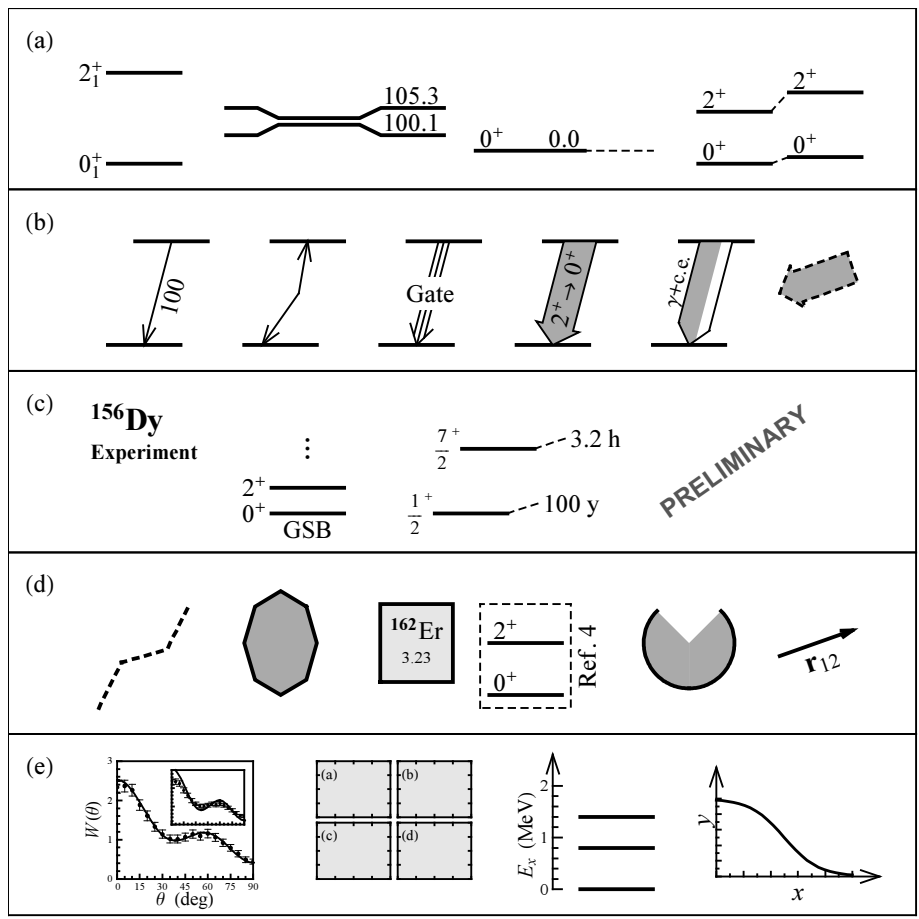}
\end{center}
\caption{Drawing objects and figure components provided by the LevelScheme
system. (a)~Levels, extension lines, and connectors.  (b)~Transition
arrows, including arrows with kinks, arrows with multiple shafts,
extended shapes with outline and fill, and split-shading conversion
electron transition arrows. (c)~Labels, either stand-alone or attached
in various ways to existing levels, making full use of Mathematica's
text-formatting capabilities.  (d)~Drawing shapes, including
polylines, polygons, boxes, filled ellipses or arcs, and arrows.
(e)~Panels and axes, with extensive capability for labeling and tick
customization.  }
\label{figgallery}
\end{figure*}

The basic drawing objects are as follows:
\begin{enumerate}
\item An energy level object (\texttt{Lev}), consisting of a line and
several attached labels.  This line may have raised or lowered end
segments (``gull wings'') to allow room for labels, as is customary in
certain types of level schemes.  There are also objects providing an
extension line to the side of a level (\texttt{ExtensionLine}) or a
connecting line between levels (\texttt{Connector}).  These are shown
in Fig.~\ref{figgallery}(a).
\item A transition arrow object (\texttt{Trans}), the endpoints of
which can be automatically attached to two given levels or can be
free-floating.  This object is quite versatile.  Labels can be
attached to the transition arrow at predefined but adjustable positions,
either aligned along the arrow or oriented as the user prefers.
The arrow can be of several different styles ranging from a simple
line to a filled polygonal arrow shape, with user control of all
dimensions.  Examples are shown in Fig.~\ref{figgallery}(b).
\item Separate label objects.  Labels may be free-standing
at coordinates specified by the user (\texttt{ManualLabel} and
\texttt{ScaledLabel}) or may be positioned in relation to an existing
level (\texttt{BandLabel} and \texttt{LevelLabel}).  See
Fig.~\ref{figgallery}(c).
\item General drawing objects.  These are essentially enhanced
versions of the basic Mathematica drawing
primatives~\cite{wolfram1999:mathematica-book-4}.  They differ from
the basic primatives in that their drawing properties are controlled
through the LevelScheme option system and in that they conveniently
combine an outline, fill, and text labels in a single object.  The
general drawing objects are an open curve or polyline
(\texttt{SchemeLine}), a closed curve or polygon
(\texttt{SchemePolygon}), a rectangle (\texttt{SchemeBox}), and a
filled circular or elliptical arc (\texttt{SchemeCircle}).  An arrow
similar to \texttt{Trans}, but meant for general drawing and
annotation tasks, is also provided (\texttt{SchemeArrow}).  These are
illustrated in Fig.~\ref{figgallery}(d).
\end{enumerate}

The user creates a figure by providing a list of such objects to the
drawing function \texttt{Scheme} (Sec.~\ref{secinfra}).  Each object's
position and appearance is governed by arguments. The most essential
positioning information is indicated through a small number of
mandatory arguments.  The other aspects of the object's positioning
and appearance are specified as options, using the standard
Mathematica
\textit{option}$\mathtt\rightarrow$\textit{value}
syntax~\cite{wolfram1999:mathematica-book-4}, with extensions
described in Sec.~\ref{secinfra}.

Each drawing object is built from up to three
distinct parts: an outline, a filled area, and attached text labels.
The labels may contain any Mathematica expression and can take full
advantage of Mathematica's extensive typesetting capabilities.
Several basic options controling the appearance of the object's parts
(\textit{e.g.},
\texttt{Thickness},
\texttt{FillColor}, and \texttt{FontSize}) are standardized across 
all object types, as are options for specifying the contents and
positions of the text labels.  Other options, such as those governing
a level's gull wings or an arrow's shape, are specific to a given
object type.  If the user does not explicitly give a value for an
option, its global default value is used.  The user can thus control
the style of all objects of a given type by changing the default
setting of the relevant option, with \texttt{SetOptions}.  For
instance, the line thicknesses of the levels in a diagram can all be
changed at once in this fashion, or the labels on these levels can be
relocated from the top to the sides.  The stylistic changes can be
applied to the whole diagram or they can be made midway through, to
affect only one portion of the diagram.

Let us consider a brief concrete example.  The function used to draw a
level object has syntax \texttt{Lev[}\textit{name}\texttt{,$x_1$,$x_2$,$E$]},
where the arguments are, respectively, a name chosen by the user, to
be used to identify the level later so that other objects such as
transition arrows can be attached to it, the left and right endpoint
$x$ coordinates, and the energy or $y$ coordinate.  The code needed to
generate the leftmost diagram in Fig.~\ref{figgallery}(b) is then
\begin{alltt}
    SetOptions[Lev, Thickness -> 2],
    Lev[lev0, 0, 1, 0], 
    Lev[lev1, 0.3, 1.3, 10], 
    Trans[lev1, lev0, LabR -> 100]
\end{alltt}
More extensive examples may be found in Appendix~\ref{appsource} and
the documentation provided through the CPC Program Library.

\section{Figure preparation tools}
\label{sectools}

Mathematica provides a powerful system for generating graphics, but it
does not by itself make available the fine formatting control
necessary for the preparation of publication-quality figures.  The
relevant figure layout and tick customization tools provided by
LevelScheme are summarized in this section.

Very often it is necessary to prepare figures with multiple parts.
These range from simple side-by-side diagrams to more complicated
possibilities, such as inset plots or rectangular
arrays of plots with shared axes [Fig.~\ref{figgallery}(e)].
LevelScheme includes a comprehensive framework for assembling
multipart figures.

The basic element of a multipart figure is a ``panel''.  This is a
rectangular region designated as a plotting window within the full
plot.  Panels can be arranged within the figure as the user pleases,
and they can arbitrarily overlap with each other.  Each panel can have
its own ranges defined for the $x$ and $y$ axes, as specified by the
user.  All objects are drawn with respect to these coordinates, as
discussed further in Sec.~\ref{secinfra}.  If the user changes the
position or size of the panel as a whole, all its contents thus move
or are rescaled accordingly, without any need for changes to the
arguments of the individual drawing objects.

A panel can be drawn with any of several ancillary components: a
frame, tick marks, tick labels, axis labels on the edges of the frame,
a panel letter, and a solid background color.  All these
characteristics are controlled by options to the basic panel
definition command \texttt{Panel}.  A panel can also simply be used
without these, as an invisible structure to aid in laying out part of
the figure.  As an alternative to axis ticks on the panel frame, the
object \texttt{SchemeAxis} allows stand-alone axes (including a line
with optional arrow head, tick marks, tick labels, and an axis label)
to be placed wherever needed within a figure
[Fig.~\ref{figgallery}(e)].

The most common arrangement of figure panels is as a
rectangular array [Fig.~\ref{figgallery}(e)].  The \texttt{Multipanel} tool greatly simplifies
the construction of such an array.  The user specifies a total region
to be covered with the panels and the numbers of rows and columns of
panels within this region, rather than calculating the position for each
panel manually.  The panels can be contiguous (shared edges) or
separated by gaps, and different rows or columns of panels can have
different heights or widths.  The axes of the panels can be
``linked'', as in many data plotting packages: The user specifies the
$y$ axis range for each row and $x$ axis range for each column, and
all panels within the row or column use this same range.  The user can
also specify the tick marks and axis labels by row and column.  (Tick
and axis labels are by default drawn only along the bottom and left
exterior edges of the array as a whole.)  Panels are automatically
labeled with panel letters, formatted as chosen by the user.  The user
has the flexibility to override any or all of these automated settings
on a panel by panel basis.

LevelScheme also supports detailed customization of tick mark
placement and formatting.  A very general function is provided to
construct sets of tick marks.  These tick marks can be used on panels,
on stand-alone axes, and on the frame of the LevelScheme figure as a
whole.  (The built-in Mathematica plot display function can generate a
default set of tick marks, but it does not give the user any control
over the tick intervals or labels.)  Linear, logarithmic, and general
nonlinear axes are supported.  
For linear axes, the user specifies the start and end points of the
coordinate range to be covered with tick marks and, optionally, the
major tick interval and minor subdivisions.  Options are used to
control the appearance of the tick marks and formatting of the labels,
or to suppress the display of certain tick marks or labels.  

Beyond these basic tick mark formatting customizations, great
flexibility is obtained through integration with the Mathematica
functional programming language.  Arbitrary nonlinear axes are
obtained by specifying various transformation functions to be applied
to the major and minor tick mark positions.  Although the tick
generation function provides basic fixed-point decimal label
formatting, the user can instead specify an arbitrary function to
generate the tick label from the tick coordinate.  The resulting label
can be any Mathematica text or symbolic expression and can involve
complicated typesetting: for instance, plots of trigonometric
functions might have tick labels written as rational multiples of the
symbol $\pi$ (see examples in the documentation provided through the
CPC Program Library).  For convenience, a predefined function is
provided for constructing logarithmic ticks with labels in the format
$b^n$.

\section{Infrastructure}
\label{secinfra}

Let us now consider the technical framework underlying this
LevelScheme figure construction process.  This includes extensions to
the Mathematica option passing scheme, definitions for handling a set
of overlaid coordinate systems, a layered drawing system, and a
mechanism for incorporating externally generated Mathematica graphics
into the figure.

Options play a major role in LevelScheme's approach to providing a
simple but flexible drawing system and are used to control most
aspects of the formatting of a level scheme.  Two conceptual extensions to
the standard Mathematica option passing scheme were needed to make
this possible.  Under the usual
option system~\cite{wolfram1999:mathematica-book-4}, default option values
for a function are defined globally in the list
\texttt{Options[}\textit{function}\texttt{]}, and these default values can be
overriden by specifying an argument
\textit{option}$\mathtt\rightarrow$\textit{value} when the function is invoked.

The user makes stylistic changes within a level scheme, affecting the
appearance of all drawing objects of a given type, by setting the
default values of relevant options, as discussed in
Sec.~\ref{secprinciples}.  However, if such a change were made to the
usual \textit{global} default value, it would have collateral effects on all
other level schemes drawn in the same Mathematica session.  To remedy
this situation, LevelScheme implements dynamic scoping of default
option value settings.  The global default values are saved before
processing of the scheme starts, and these original values are
restored after processing finishes, so any changes made within a level
scheme are confined to that scheme.

It is also sometimes convenient to have changes
to certain basic stylistic options, such as the font size or line
thickness, affect \textit{all} objects in a drawing, not just those of
a given type.  To facilitate this, the concept of inheritance, from
object oriented programming, has been applied to the Mathematica
option system.  All LevelScheme drawing objects are all defined to be
child objects of a common parent object \texttt{SchemeObject}.  The values of the
basic outline, fill, and text style options (Sec.~\ref{secobjects})
for these objects are by default taken, or inherited, from values set for
\texttt{SchemeObject}.  (Further details are given 
in Ref.~\cite{mathsourcecontribs}.)

\begin{figure}
\begin{center}
\includegraphics*[width=0.8\hsize]{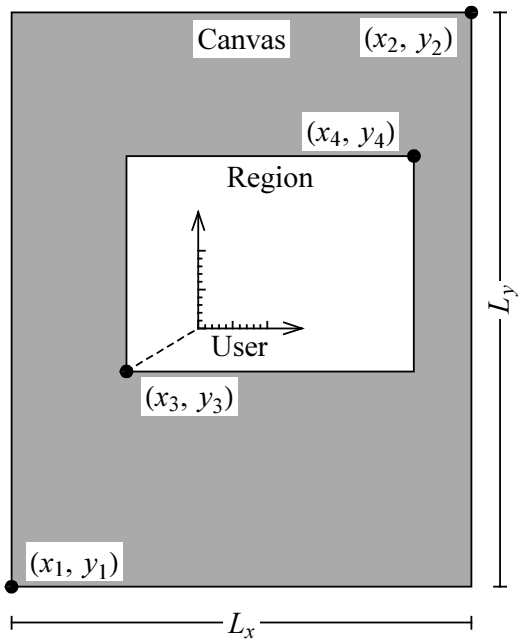}
\end{center}
\caption{Overlaid coordinate systems maintained by LevelScheme:
absolute (or physical), canvas, region, and user systems.
}
\label{figcoords}
\end{figure}

Several different, complementary coordinate systems are needed to
describe points within a LevelScheme figure (see
Fig.~\ref{figcoords}).  The Mathematica graphics display functions
only recognize a single coordinate system, which runs from 
user-specified coordinates $(x_1,y_1)$ at the lower left corner of the figure to
$(x_2,y_2)$ at the upper right corner.  We refer to these coordinates
as the ``canvas coordinates''.  All graphics coordinates must
ultimately be expressed in terms of these for display by Mathematica.%

However, many drawing tasks require the calculation of physical
distances as they will appear on the output page.  Angles are not
preserved under an unequal rescaling of the horizontal and vertical
axes.  Thus, for instance, construction of the right angled corners in
arrows [Fig.~\ref{figgallery}(b)] requires knowledge of physical
horizontal and vertical distances rather than just $x$ and $y$
coordinate distances.  Since the physical dimensions $L_x$ and $L_y$
of the plotting region are known (Fig.~\ref{figcoords}), 
``absolute coordinates'' measuring the physical position of a point
within the figure can easily be related to the canvas coordinates.

For constructing multipart figures (Sec.~\ref{sectools}), a smaller
rectangular plotting region is designated within the full figure, with
new ranges defined for the $x$ and $y$ axes within this region.  The
resulting ``region coordinates'' are fully determined if the
coordinate values of the corners of the region, $(x_3,y_3)$ and
$(x_4,y_4)$, are specified in both the canvas and region coordinate
systems.  These region coordinates are the basic coordinates defined
within a plot panel.  However, it is also convenient for the user to
be able to arbitrarily shift portions of a diagram collectively,
without individually modifying the coordinates specified for all the
objects involved.  This is accomplished by introducing ``user
coordinates'', which have a user-defined offset and scale relative to
the region coordinates.  (For instance, the individual diagrams were
arranged from left to right within each panel of Fig.~\ref{figgallery}
by defining a different horizontal user coordinate offset before
drawing the objects for each one.)  The user specifies all coordinates
for LevelScheme objects in the user coordinate system.  Initially, the
canvas, region, and user coordinate systems for a figure are
identical.  Each time the user redefines the coordinates,
\textit{e.g.}, by defining a panel or introducing a user coordinate
offset, the affine transformation coefficients relating the various coordinate systems are recalculated
and stored.  

A final important component of the drawing infrastructure is the
layered drawing system.  Each drawing element is assigned to a
numbered layer.  Those assigned to lower-numbered layers (background)
are rendered before, and thus might be hidden by, objects assigned to
higher-numbered layers (foreground).  The layering system is essential
for preventing text labels from being hidden by other drawing elements
in dense level schemes.  By default, outlines and fills are assigned
to layer 1, ``white-out'' boxes behind text labels are in layer 2, and
the text itself is in layer 3.  With this layering system, white-out
boxes hide any lines or fills behind them, but they do not block
neighboring text, making possible closely-spaced transition labels,
such as are needed in decay schemes (see
Sec.~\ref{secexamples} for examples).
\nocite{sutherland1974:polygon-clipping}
\afterpage{
\clearpage
\begin{figure}[p]
\begin{center}
\includegraphics*[width=\hsize]{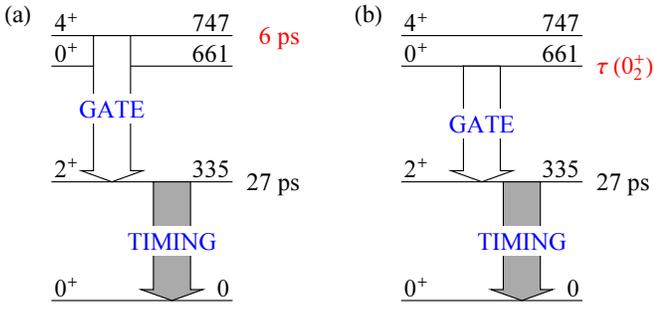}
\end{center}
\caption{A simple level scheme, typical of
those used in presentations, with filled polygonal transition arrows.
Figure from Ref.~\cite{caprio2003:diss}.  }
\label{fig154dycascades}
\end{figure}
\begin{figure}[p]
\begin{center}
\includegraphics*[width=0.9\hsize]{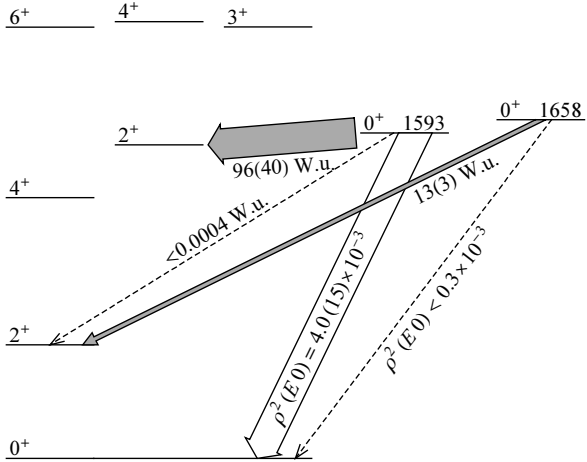}
\end{center}
\caption{Level scheme illustrating various arrow styles.  Figure from
Ref.~\cite{caprio2003:diss}.  }
\label{fig102pd0plus}
\end{figure}
\begin{figure}[p]
\begin{center}
\includegraphics*[width=0.7\hsize]{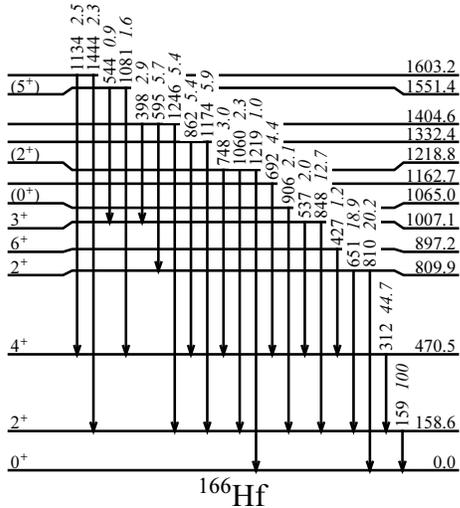}
\end{center}
\caption{A decay scheme in the classic style, created using the
LevelScheme decay scheme generation tools.  Figure from
Ref.~\cite{mccutchan2005:166hf-beta}.  }
\label{fig166hfbeta}
\end{figure}
\begin{figure}[p]
\begin{center}
\includegraphics*[width=\hsize]{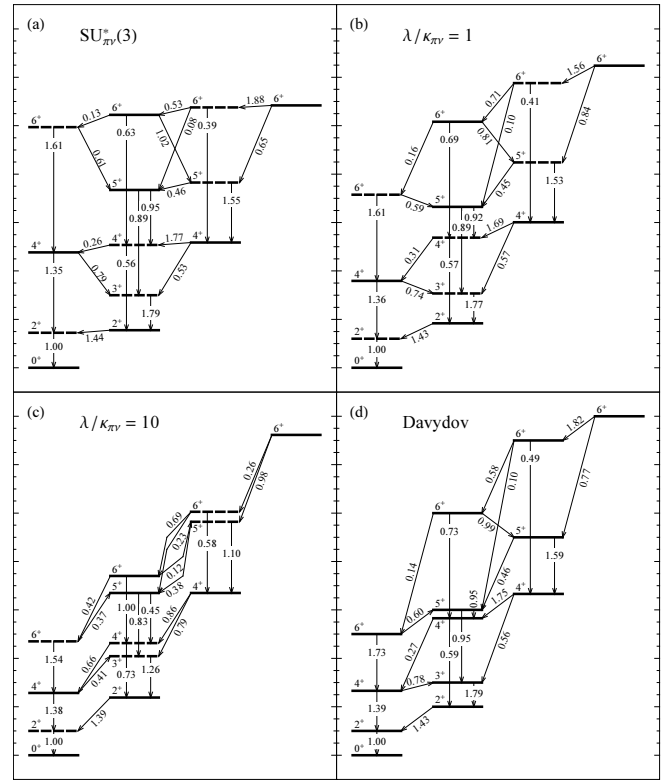}
\end{center}
\caption{A multipanel level scheme.  Figure from
Ref.~\cite{caprio2005:ibmpn2}.  }
\label{figtriaxschemes}
\end{figure}
\begin{figure}[p]
\begin{center}
\includegraphics*[width=\hsize]{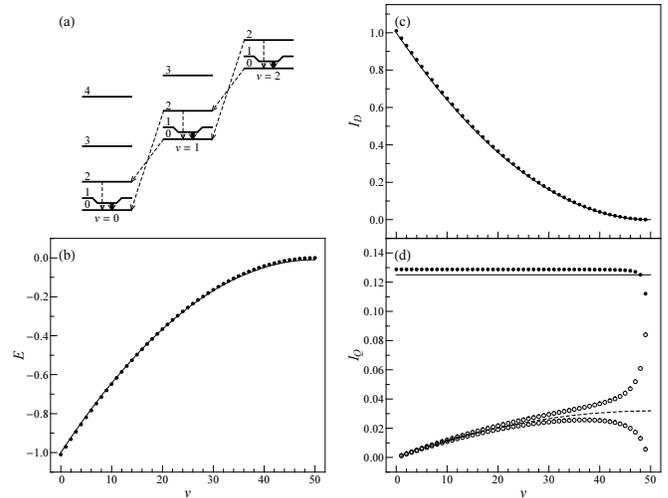}
\end{center}
\caption{A multipanel figure combining a level scheme diagram with
function and data plots.  Figure from
Ref.~\cite{caprio2005:coherent}.  }
\label{figcombo}
\end{figure}
\begin{figure*}[p]
\begin{center}
\includegraphics*[width=0.8\hsize]{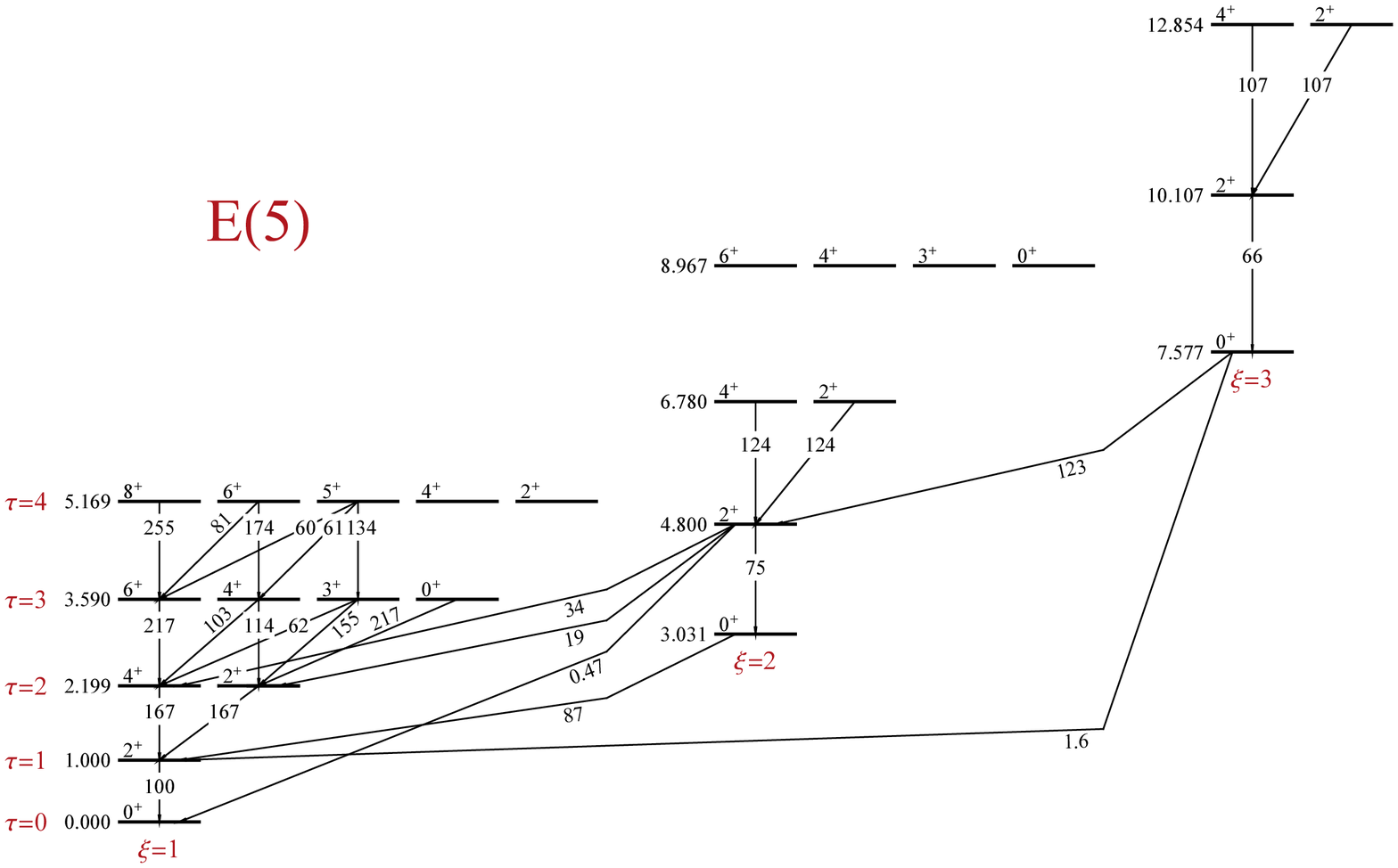}
\end{center}
\caption{A large reference level scheme, with bent transition arrows
and extensive use of annotation labels for families of levels.  Figure from
Ref.~\cite{caprio2003:diss}.  }
\label{fige5reference}
\end{figure*}
\begin{figure*}[p]
\begin{center}
\includegraphics*[width=0.48\hsize]{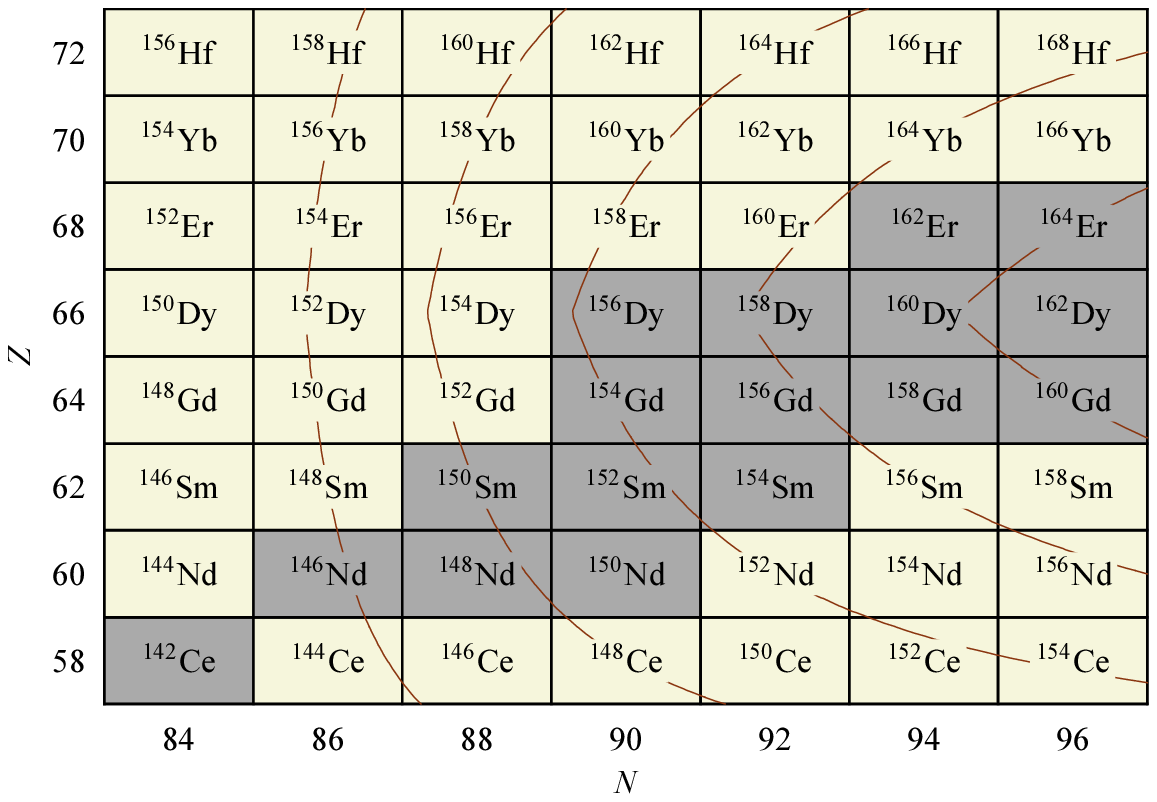}
\hfill
\includegraphics*[width=0.48\hsize]{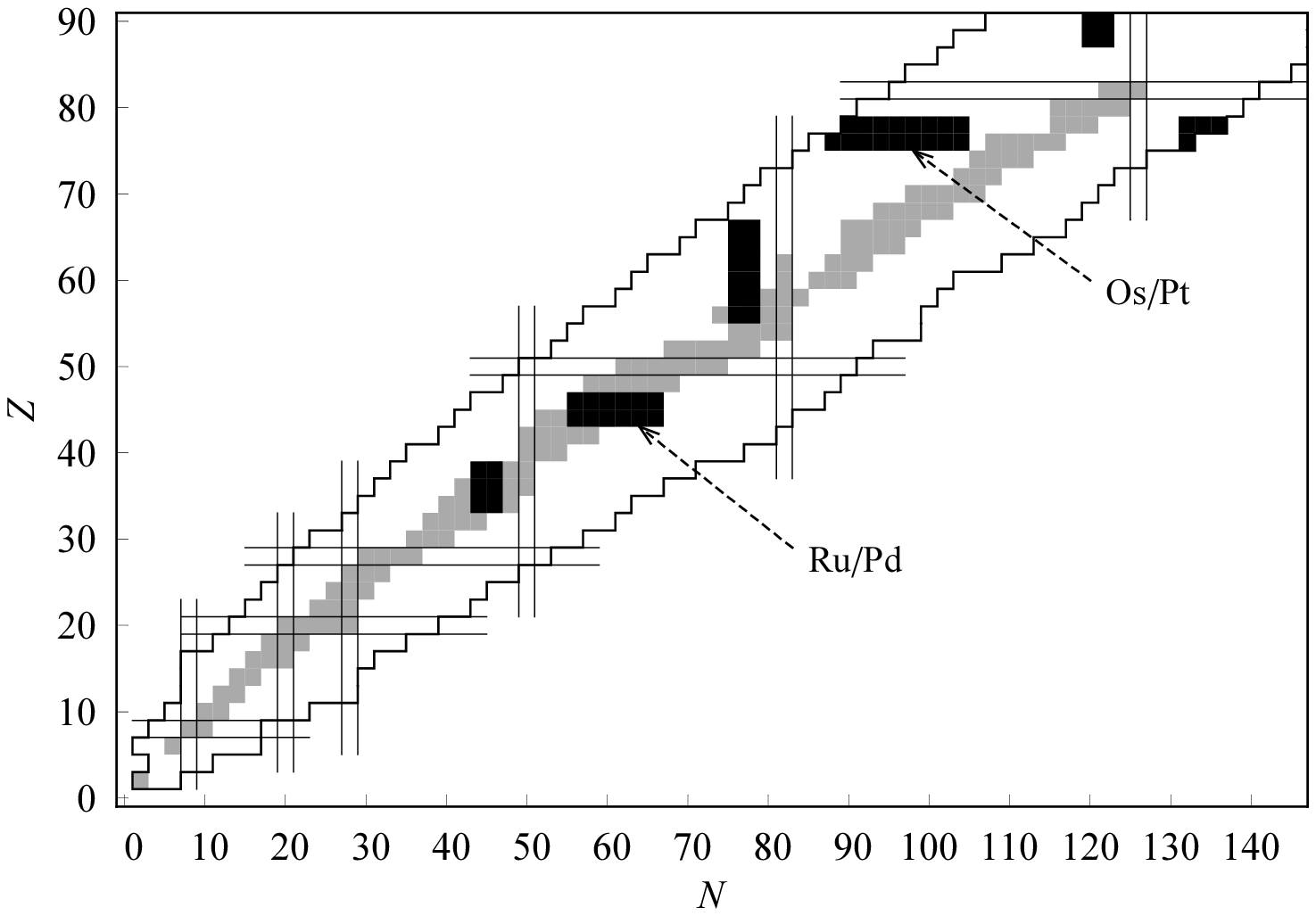}
\end{center}
\caption{
Nuclear charts created using LevelScheme drawing
objects and the Mathematica chemical elements database (see text).
Right panel adapted from Ref.~\cite{caprio2005:ibmpn-nmp04}.
}
\label{fignchart}
\end{figure*}
\clearpage
}

Arbitrary Mathematica graphics, including the output of the
Mathematica plotting routines, can be incorporated into a LevelScheme
figure.  Mathematica represents graphical output as a
\texttt{Graphics} object containing a list of
``primatives''~\cite{wolfram1999:mathematica-book-4}, which are either
drawing elements (polylines, polygons, points, \textit{etc.})  or
directives affecting the style in which the following drawing elements
are rendered (color, thickness, dashing).  For inclusion in a
LevelScheme figure, such graphics must be ``wrapped'' in a
LevelScheme
\texttt{RawGraphics} object, which extracts the primatives from the 
\texttt{Graphics} object and carries out
several manipulations on them.  It first transforms all the
coordinates contained in the graphics primatives from user coordinates
to canvas coordinates, thereby moving and scaling the graphics to
appear according to the user's current choice of coordinate system.
(The graphics may contain coordinates given in the special
\texttt{Offset} and \texttt{Scaled} notations defined by
Mathematica.)  It then clips all polylines and polygons to the current
plot region, using a standard
algorithm~\cite{sutherland1974:polygon-clipping}.  Finally, it
supplies default drawing style primatives and a layer number for the
graphics, for use by the layered drawing system.

The graphical content of a LevelScheme figure is assembled by the
function \texttt{Scheme}.  The user provides a list of LevelScheme drawing
objects to \texttt{Scheme}.  Intermixed with these objects may be
commands, such as \texttt{SetOptions} or \texttt{Panel}, which affect
the appearance of the following objects.
\texttt{Scheme} also accepts options governing the overall figure
properties (plot range, dimensions, frame labels and ticks,
\textit{etc.}) much like the usual Mathematica plotting functions.
The final output of \texttt{Scheme} is a \texttt{Graphics} object.

We can now summarize the steps carried out by \texttt{Scheme} in
creating a figure.  The list of objects passed to scheme is initially
``held'' in unevaluated form, contrary to the usual Mathematica
symbolic evaluation rules, to prevent premature evaluation of the
coordinate arithmetic expressions and any
\texttt{SetOptions} calls it contains.  \texttt{Scheme} first initializes the variables
controlling the overlayed coordinate systems, based upon the plot
range and image dimensions specified by the user.
\texttt{Scheme} then evaluates the list of objects, 
dynamically scoping the drawing object default option values as
described above.  Evaluation of the objects relies upon three
standardized functions for generating outlines, fills, and text
labels.  These functions in turn each return a structure
containing three parts: a list of graphics style primatives based upon
the LevelScheme style option values (Sec.~\ref{secobjects}), a list of
primatives for the drawing elements, and a layer number.  \texttt{Scheme} sorts
the resulting structures in order of ascending layer number, to insure
that drawing elements in foreground layers are rendered after
those in background layers, and then discards the layer
information, extracting just the graphics primatives.  Since many
consecutive objects typically have the same drawing style properties,
\texttt{Scheme} achieves considerably more compact output by stripping 
redundant style primatives from the list.
\texttt{Scheme} also constructs frame and axis directives for the whole
scheme, which are combined with the graphics primative list to make a
Mathematica \texttt{Graphics} object.  

\section{Graphical output examples}
\label{secexamples}

Several examples of figures generated using the LevelScheme system are
shown in this section.  These figures, mainly level scheme diagrams,
have been chosen to demonstrate several of the capabilities discussed
in the earlier sections and to illustrate the variety of drawing
styles possible.

Figs.~\ref{fig154dycascades} and~\ref{fig102pd0plus} are basic level
schemes, as might be encountered in presentations.  Note the various
arrow styles and arrow label positions.  The complete code used to
generate Fig.~\ref{fig154dycascades} is given in
Appendix~\ref{appsource}.

Fig.~\ref{fig166hfbeta} is in the classic style for a decay scheme,
showing $\gamma$-ray transitions among nuclear levels populated
following $\alpha$ or $\beta$ decay.  LevelScheme provides tools which
automate the positioning of the equally-spaced vertical transition
arrows in such decay schemes.  Note the gull wings on the levels and
the white-out boxes behind the transition labels.  The layered drawing
system discussed in Sec.~\ref{secinfra} prevents these boxes from
obstructing neighboring labels.

Fig.~\ref{figtriaxschemes} is an example of a multipanel figure
involving only level schemes.  Fig.~\ref{figcombo} combines a level
scheme diagram with function and data plots.

Fig.~\ref{fige5reference} is a larger-scale scheme, involving bent
arrows.  Heavy use is made of annotation labels beside or below
levels.  In such a scheme, user coordinate offsets simplify adjustment
of the horizontal spacing between the different families of levels.

Fig.~\ref{fignchart} provides an illustration of the use of
LevelScheme drawing tools for technical diagrams other than level
schemes.  Mathematica's programming language can be used to automate the
construction of complex diagrams containing large numbers of drawing
objects.  Nuclear charts are simple to create using the
\texttt{SchemeBox} drawing object, which provides the outline, fill,
and labels for each cell.  The entire array of cells can be
constructed by using a single Mathematica \texttt{Table} construct to
iterate over proton and neutron numbers.  (Mathematica's chemical
elements database provides the element symbols and information on
which nuclides are stable.) The resulting chart can be overlaid with a
Mathematica contour plot, as in Fig.~\ref{fignchart} (left), where the
LevelScheme layered drawing system has been used to ensure that labels
appear in front of the contour lines.  Information read from external
data files can be superposed as well, in the form of cell colors, cell
text contents, or region boundary lines, as in Fig.~\ref{fignchart}
(right).

\section{Conclusion}
\label{secconcl}

The LevelScheme system for Mathematica provides a flexible system for
the construction of level energy diagrams, automating many tedious aspects
of the preparation while allowing for extensive manual fine-tuning and
customization.  The general figure preparation tools and
infrastructure developed for this purpose also have broad
applicability to the preparation of publication-quality multipart
figures with Mathematica, incorporating diagrams, mathematical plots,
and data plots.

\begin{acknowledgments}
Discussions with and feedback from E.~A.~McCutchan, M.~Babilon,
N.~V.~Zamfir, and M.~Fetea are gratefully acknowledged.
Fig.~\ref{fig166hfbeta} was provided by E.~A.~McCutchan.  This work
was supported by the US DOE under grant DE-FG02-91ER-40608.
\end{acknowledgments}

\appendix
\onecolumngrid
\section{Figure source code}
\label{appsource}

This appendix contains the Mathematica code used to generate
Fig.~\ref{fig154dycascades} with the LevelScheme figure preparation
system.


\begin{alltt}
    Scheme[\{
      
      (* set default styles for level scheme objects *)
      
      SetOptions[SchemeObject, FontSize -> 20];
      SetOptions[Lev, NudgeL -> 1, NudgeR -> 1, LabR -> Automatic],
      SetOptions[Trans, ArrowType -> ShapeArrow, HeadLength -> 9, HeadLip -> 10, Width -> 30,
        FontColor -> Blue, OrientationC -> Horizontal, BackgroundC -> Automatic],
      SetOptions[LevelLabel, Gap -> 10],
      
      (* draw cascade from 4+ level *)
      
      SetOrigin[0],
      ManualLabel[\{-.4, 850\}, "(a)", Offset -> \{-1, 1\}],
      
      Lev[lev0, 0, 2, "0", LabL -> LabelJP[0]],
      Lev[lev335, 0, 2, "335", LabL -> LabelJP[2]],
      LevelLabel[lev335, Right, "27 ps"],
      Lev[lev661, 0, 2, "661", LabL -> LabelJP[0]],
      Lev[lev747, 0, 2, "747", LabL -> LabelJP[4]],
      LevelLabel[lev747, Right, "  6 ps", FontColor -> Red],
      Trans[lev335, 1.3, lev0, Automatic, FillColor -> LightGray, LabC -> "TIMING"],
      Trans[lev747, .7, lev335, Automatic, FillColor -> White, LabC -> "GATE"],
      
      (* draw cascade from 0+ level *)
      
      SetOrigin[3.5],
      ManualLabel[\{-.4, 850\}, "(b)", Offset -> \{-1, 1\}],
      
      Lev[lev0, 0, 2, "0", LabL -> LabelJP[0]],
      Lev[lev335, 0, 2, "335", LabL -> LabelJP[2]],
      LevelLabel[lev335, Right, "27 ps"],
      Lev[lev661, 0, 2, "661", LabL -> LabelJP[0]],
      LevelLabel[lev661, Right, 
        RowBox[\{"\begin{math}\mathtt\tau\end{math}(", hspace[-0.2], LabelJiP[0, 2], ")"\}], 
        FontColor -> Red],
      Lev[lev747, 0, 2, "747", LabL -> LabelJP[4]],
      Trans[lev335, 1.3, lev0, Automatic, FillColor -> LightGray, LabC -> "TIMING"],
      Trans[lev661, .9, lev335, Automatic, FillColor -> White, LabC -> "GATE"],
      
      \},
    PlotRange -> \{\{-.8, 6.3\}, \{-100, 900\}\}, ImageSize -> 72*\{8, 4\}
    ];
\end{alltt}
\twocolumngrid

\vfil


\providecommand{\ELSEVIER}{}
\ELSEVIER\newcommand{\identity}[1]{{#1}}



\begin{thebibliography}{13}
\expandafter\ifx\csname natexlab\endcsname\relax\def\natexlab#1{#1}\fi
\expandafter\ifx\csname bibnamefont\endcsname\relax
  \def\bibnamefont#1{#1}\fi
\expandafter\ifx\csname bibfnamefont\endcsname\relax
  \def\bibfnamefont#1{#1}\fi
\expandafter\ifx\csname citenamefont\endcsname\relax
  \def\citenamefont#1{#1}\fi
\expandafter\ifx\csname url\endcsname\relax
  \def\url#1{\texttt{#1}}\fi
\expandafter\ifx\csname urlprefix\endcsname\relax\def\urlprefix{URL }\fi
\providecommand{\bibinfo}[2]{#2}
\providecommand{\eprint}[2][]{\url{#2}}

\bibitem{wolfram1999:mathematica-book-4}
\bibinfo{author}{\bibfnamefont{S.}~\bibnamefont{Wolfram}},
  \emph{\bibinfo{title}{The Mathematica Book}}, \bibinfo{edition}{4th} ed.
  (\bibinfo{publisher}{Wolfram Media/Cambridge University Press},
  \bibinfo{year}{1999}).

\bibitem{wolfram1999:mathematica-4}
\bibinfo{author}{\bibnamefont{{Wolfram Research, Inc.}}},
  \emph{\bibinfo{title}{Mathematica 4}} (\bibinfo{address}{Champaign,
  Illinois}, \bibinfo{year}{1999}).

\bibitem{hahn2001:feynarts}
\bibinfo{author}{\bibfnamefont{T.}~\bibnamefont{Hahn}},
  \bibinfo{journal}{Comput. Phys. Commun.} 111 (1998) 217.

\bibitem{radford1995:radware}
\bibinfo{author}{\bibfnamefont{D.~C.} \bibnamefont{Radford}},
  \bibinfo{journal}{Nucl. Instrum. Methods A} 361 (1995) 297.

\bibitem{dunford2003:ensdat}
\bibinfo{author}{\bibfnamefont{C.~L.} \bibnamefont{Dunford}} \bibnamefont{and}
  \bibinfo{author}{\bibfnamefont{R.~R.} \bibnamefont{Kinsey}},
  \bibinfo{note}{computer code ENSDAT (unpublished)}.

\bibitem{levelschemehome}
\bibinfo{note}{Updates and further information may be obtained through the
  LevelScheme home page
  (\texttt{http://\linebreak[0]wnsl.physics.yale.edu/\linebreak[0]levelscheme}%
).}

\bibitem{mathsourcecontribs}
\bibinfo{note}{The BlockOptions, CustomTicks, ForEach, and InheritOptions
  components of the LevelScheme system have previously been made available
  through the Mathematica Information Center's \textit{MathSource} code library
  (\texttt{http://\linebreak[0]library.wolfram.com/\linebreak[0]infocenter}),
  where further technical documentation may be found.}

\bibitem{sutherland1974:polygon-clipping}
\bibinfo{author}{\bibfnamefont{I.~E.} \bibnamefont{Sutherland}}
  \bibnamefont{and} \bibinfo{author}{\bibfnamefont{G.~W.}
  \bibnamefont{Hodgman}}, \bibinfo{journal}{Commun. ACM} 17 (1974) 132.

\bibitem{caprio2003:diss}
\bibinfo{author}{\bibfnamefont{M.~A.} \bibnamefont{Caprio}}, Ph.D. thesis,
  \bibinfo{school}{Yale University} (\bibinfo{year}{2003}),
  \eprint{arXiv:\linebreak[0]nucl-ex/\linebreak[0]0502004}.

\bibitem{mccutchan2005:166hf-beta}
\bibinfo{author}{\bibfnamefont{E.~A.} \bibnamefont{McCutchan}},
  \bibinfo{author}{\bibfnamefont{N.~V.} \bibnamefont{Zamfir}},
  \bibinfo{author}{\bibfnamefont{R.~F.} \bibnamefont{Casten}},
  \bibinfo{author}{\bibfnamefont{M.~A.} \bibnamefont{Caprio}},
  \bibinfo{author}{\bibfnamefont{H.}~\bibnamefont{Ai}},
  \bibinfo{author}{\bibfnamefont{H.}~\bibnamefont{Amro}},
  \bibinfo{author}{\bibfnamefont{C.~W.} \bibnamefont{Beausang}},
  \bibinfo{author}{\bibfnamefont{A.~A.} \bibnamefont{Hecht}},
  \bibinfo{author}{\bibfnamefont{D.~A.} \bibnamefont{Meyer}}, \bibnamefont{and}
  \bibinfo{author}{\bibfnamefont{J.~J.} \bibnamefont{Ressler}},
  \bibinfo{journal}{Phys. Rev. C} 71 (2005) 024309.

\bibitem{caprio2005:ibmpn2}
\bibinfo{author}{\bibfnamefont{M.~A.} \bibnamefont{Caprio}} \bibnamefont{and}
  \bibinfo{author}{\bibfnamefont{F.}~\bibnamefont{Iachello}},
  \bibinfo{journal}{Ann. Phys. (N.Y.)} 318 (2005) 454.

\bibitem{caprio2005:coherent}
\bibinfo{author}{\bibfnamefont{M.~A.} \bibnamefont{Caprio}},
  \bibinfo{journal}{J. Phys. A} 38 (2005) 6385.

\bibitem{caprio2005:ibmpn-nmp04}
\bibinfo{author}{\bibfnamefont{M.~A.} \bibnamefont{Caprio}}, in
  \emph{\bibinfo{booktitle}{Nuclei and Mesoscopic Physics}}, edited by
  \bibinfo{editor}{\bibfnamefont{V.}~\bibnamefont{Zelevinsky}},
  \bibinfo{series}{AIP Conf. Proc.} No. \bibinfo{number}{777}
  (\bibinfo{publisher}{AIP}, \bibinfo{address}{Melville, New York},
  \bibinfo{year}{2005}), p. \bibinfo{pages}{199}.

\end{thebibliography}
\end{document}